\newcommand{\Tr}{\mathop{\mathrm{Tr}}\nolimits}

\newcommand{\op}[1]{\hat{#1}}

\documentclass[pra,twocolumn,showpacs,amsmath,amssymb,superscriptaddress,eqsecnum]{revtex4-1}
\usepackage{graphicx,bm,color,mathptmx,hyperref,dcolumn} 
\begin{document}

\title{Multipolar hierarchy of efficient quantum polarization measures}

\author{P.~de la Hoz} 
\affiliation{Departamento de \'Optica, 
Facultad  de F\'{\i}sica, Universidad Complutense, 
28040~Madrid, Spain}

\author{A.~B.~Klimov} 
\affiliation{Departamento de F\'{\i}sica,
  Universidad de Guadalajara, 44420~Guadalajara, Jalisco, Mexico}

\author{G.~Bj\"{o}rk} 
\affiliation{Department of Applied Physics,
  Royal Institute of Technology (KTH), AlbaNova University Center,
  SE-106 91 Stockholm, Sweden}

\author{Y.-H. Kim} 
\affiliation{Department of Physics, 
Pohang University of Science and  Technology (POSTECH), 
Pohang, 790-784, Korea}

\author{C.~M\"{u}ller} 
\affiliation{Max-Planck-Institut f\"ur die Physik des Lichts, 
 G\"{u}nther-Scharowsky-Stra{\ss}e 1, Bau 24,
  91058 Erlangen, Germany} 
\affiliation{Department f\"{u}r Physik,
  Universit\"{a}t Erlangen-N\"{u}rnberg, Staudtstra{\ss}e 7, Bau 2,
  91058 Erlangen, Germany}

\author{Ch.~Marquardt} 
\affiliation{Max-Planck-Institut f\"ur die Physik des Lichts, 
 G\"{u}nther-Scharowsky-Stra{\ss}e 1, Bau 24,
  91058 Erlangen, Germany} 
\affiliation{Department f\"{u}r Physik,
  Universit\"{a}t Erlangen-N\"{u}rnberg, Staudtstra{\ss}e 7, Bau 2,
  91058 Erlangen, Germany}

\author{G.~Leuchs} 
\affiliation{Max-Planck-Institut f\"ur die Physik des Lichts, 
 G\"{u}nther-Scharowsky-Stra{\ss}e 1, Bau 24,
  91058 Erlangen, Germany} 
\affiliation{Department f\"{u}r Physik,
  Universit\"{a}t Erlangen-N\"{u}rnberg, Staudtstra{\ss}e 7, Bau 2,
  91058 Erlangen, Germany}

\author{L.~L.~S\'anchez-Soto} 
\affiliation{Departamento de \'Optica,
  Facultad de F\'{\i}sica, Universidad Complutense, 
28040~Madrid,  Spain} 
\affiliation{Max-Planck-Institut f\"ur die Physik des Lichts, 
 G\"{u}nther-Scharowsky-Stra{\ss}e 1, Bau 24,
  91058 Erlangen, Germany} 
\affiliation{Department f\"{u}r Physik,
  Universit\"{a}t Erlangen-N\"{u}rnberg, Staudtstra{\ss}e 7, Bau 2,
  91058 Erlangen, Germany}


\date{\today}

\begin{abstract}
  We advocate a simple multipole expansion of the polarization density
  matrix. The resulting multipoles appear as successive moments of the
  Stokes variables and can be obtained from feasible measurements. In
  terms of these multipoles, we construct a whole hierarchy of measures
  that accurately assess higher-order polarization fluctuations.
\end{abstract}

\maketitle

\section{Introduction} 

The standard notion of polarization comes
from the treatment of light as a beam. This hints at a well-defined
direction of propagation, and thus at a specific transverse plane,
wherein the tip of the electric field describes an ellipse. This
polarization ellipse can be elegantly visualized by using the
Poincar\'e sphere and is determined by the çStokes parameters; the
degree of polarization being simply the length of the Stokes
vector~\cite{Brosseau:1998lr}.

This geometric representation not only provides remarkable insight,
but also greatly simplifies otherwise complex problems and, as a
result, has become an indisputable tool to deal with polarization
phenomena. However, the necessity of addressing new issues, such as
highly nonparaxial fields~\cite{Nicholls:1972zr}, narrow-band imaging
systems~\cite{Pohl:1984fk}, and the recognition of associated
propagation questions~\cite{Petruccelli:2010ly}, has brought about
significant modifications of this simple classical
picture~\cite{Samson:1973ve,Barakat:1977qf,Setala:2002oq,Luis:2005kl,
  Ellis:2005bh,Refregier:2006fj,Dennis:2007tg,Sheppard:2011hc,Qian:2011kx}.

In the quantum domain, the classical setting can be immediately
mimicked in terms of the Stokes operators, which can be obtained from
the Stokes parameters by quantizing the field
amplitudes~\cite{Luis:2000ys}.  However, the appearance of hurdles
such as hidden polarization~\cite{Klyshko:1992wd}, the fact that the
Poincar\'{e} sphere cannot accommodate photon-number
fluctuations~\cite{Muller:2012ys}, and the difficulties in defining
polarization properties of two-photon entangled
fields~\cite{Jaeger:2003ij}, to cite only a few examples, show that
the resulting theory is insufficient.

The root of these difficulties can be traced to the fact that
classical polarization is chiefly built on first-order moments of the
Stokes variables, whereas higher-order moments can play a major role
for quantum fields. Polarization
squeezing~\cite{Chirkin:1993dz,*Korolkova:2002fu},  a nonclassical
effect that is actually defined only by the variances of the Stokes
operators, illustrates that point in the most clear way.

Nowadays, there is a general consensus in that a full understanding of
the subtle polarization effects arising in the realm of the quantum
world would require a characterization of higher-order
polarization fluctuations, as it happens in coherence theory, where
one needs, in general, a hierarchy of correlation functions.  Some
results along these lines have already been reported, but either they
use magnitudes difficult to determine in practice, such as
distances~\cite{Klimov:2005kl}, generalized
visibilities~\cite{Bjork:2000kx,Sehat:2005wd,Iskhakov:2011uq,Singh:2013uq},
and central moments~\cite{Bjork:2012zr}, or they go only up to second
order~\cite{Klimov:2010uq,Singh:2013ly}, and the pertinent extensions
are difficult to discern.

In this paper, we propose a systematic and feasible solution to such a
fundamental and longstanding problem. To that end, we resort to a
multipole expansion of the density matrix that naturally sorts
successive moments of the Stokes variables. The dipole term, being
just the first-order moment, can be identified with the classical
picture, while the other multipoles account for higher-order
moments.  The probability distribution for these multipoles
provides thus a complete information about the polarization properties
of any state; in terms of it we propose a suitable measure for the
quantitative assessment of those fluctuations.

\section{Setting the scenario} 

Throughout, we assume a monochromatic quantum field specified by two
operators $\op{a}_{H}$ and $\op{a}_{V}$, representing the complex
amplitudes in two linearly polarized orthogonal modes, which we denote as
horizontal ($H$) and vertical ($V$), respectively. The Stokes
operators can be concisely defined as
\begin{equation}
  \label{eq:1}
  \op{S}_{\mu} = \textstyle{\frac{1}{2}}
  \begin{pmatrix} 
    \op{a}_{H}^{\dagger} & \op{a}_{V}^{\dagger}
  \end{pmatrix}
  \sigma_{\mu}
  \begin{pmatrix}
    \op{a}_{H} \\
    \op{a}_{V}
  \end{pmatrix} \, ,
\end{equation}
the subscript $\dagger$ denoting the Hermitian adjoint. The Greek
index $\mu$ runs from 0 to 3, with $\sigma_{0} = \openone$ and $\{
\sigma_{k} \} $ ($k = 1, 2, 3$) are the Pauli matrices.

Note carefully that $\op{S}_{0} = \op{N}/2$, where $\op{N} =
\op{a}_{H}^{\dagger} \op{a}_{H} + \op{a}_{V}^{\dagger} \op{a}_{V}$ is
the operator for the total number of photons. On the other hand, with
our definition the average of $\mathbf{\op{S}} = (\op{S}_{1},
\op{S}_{2}, \op{S}_{3})$ differs by a factor of 1/2 from the classical
Stokes vector~\cite{Luis:2000ys}. However, in this way $\{
\op{S}_{k} \}$ satisfy the commutation relations of the su(2) algebra
\begin{equation}
[ \op{S}_{k}, \op{S}_\ell] = i \epsilon_{k \ell m} \, \op{S}_{m} \, ,
\end{equation} 
where $\epsilon_{k \ell m}$ is the Levi-Civita fully antisymmetric
tensor.  This noncommutability precludes the simultaneous exact
measurement of the physical quantities they represent, which can be
formulated quantitatively  by the uncertainty relation
\begin{equation}
\label{eq:unrel}
  \Delta^{2} \op{\mathbf{S}}  = 
  \Delta^{2} \op{S}_{1}  + \Delta^{2} \op{S}_{2}  + \Delta^{2} \op{S}_{3}  
  \geq \textstyle\frac{1}{2}  \langle \op{N} \rangle  \, ,
\end{equation}
where the variances are given by $\Delta^{2} \op{S}_{i} = \langle
\op{S}_{i}^{2} \rangle - \langle \op{S}_{i} \rangle^{2}$.  In other
words, the electric vector of a monochromatic quantum field never
traces a definite ellipse.

In classical optics, the states of definite polarization are specified
by $\langle \op{\mathbf{S}} \rangle^{2} = \langle \op{S}_{0}
\rangle^{2}$ and the average intensity is a well-defined quantity.  In
the three-dimensional space of the Stokes parameters this defines a
sphere with radius equal to the intensity: the Poincar\'e sphere.  In
contradistinction, in quantum optics we have that $\op{\mathbf{S}}^{2}
= \op{S}_{0} (\op{S}_{0}+ \op{\openone})$. As fluctuations in
the number of photons are, in general, unavoidable, we are forced to
work with a full three-dimensional Poincar\'e space that can be
regarded as a set of nested spheres with radii proportional to the
different photon numbers that contribute to the state.

The Hilbert space $\mathcal{H}$ of these fields is spanned by the Fock
states $ \{ |n_{H}, n_{V} \rangle \}$ for both polarization
modes. However, since $[ \op{N}, \op{\mathbf{S}} ] = 0$, each subspace
with a fixed number of photons $N$ (i.e., fixed spin $S \equiv S_{0} =
N/2$) must be handled separately. In other words, in the previous
onion-like picture of the Poincar\'e space, each shell has to be
addressed independently.  This can be underlined if we employ the
relabelling
\begin{equation}
  | S, m \rangle
  \equiv 
  | n_{H} = S + m, n_{V} = S - m \rangle \, . 
\end{equation}
In this angular momentum basis, $S = N/2$, $m = (n_{H}~-~n_{V})/2$,
and, for each $S$, $m$ runs from $-S$ to $S$.  This can be seen as the
basis of common eigenstates of $\{ \op{\mathbf{S}}^{2}, \op{S}_{3}\}$ and these
states span a $(2S+1)$-dimensional subspace wherein $\op{\mathbf{S}}$
acts in the standard way.

\section{The polarization sector and the multipole expansion} 

From the previous discussion, it is clear that the moments of any
energy-preserving observable (such as $\op{\mathbf{S}}$) do not depend
on the coherences between different subspaces.  The only accessible
information from any state described by the density matrix
$\op{\varrho}$ is thus its polarization sector~\cite{Raymer:2000zt},
which is given by the block-diagonal form
\begin{equation}
  \op{\varrho}_{\mathrm{pol}} = \bigoplus_{S} P_{S} \
  \op{\varrho}^{(S)} \, , 
\end{equation}
where $P_{S}$ is the photon-number distribution ($S$ takes on the values
0, 1/2, 1, $\ldots$) and $P_{S} \, \op{\varrho}^{(S)}$ is the reduced
density matrix in the subspace with spin $S$. Any $\op{\varrho}$ and
its associated block-diagonal form $ \op{\varrho}_{\mathrm{pol}} $
cannot be distinguished in polarization measurements and so, accordingly,
we drop henceforth the subscript pol.  This is consistent with the
fact that polarization and intensity are, in principle, separate
concepts: in classical optics the form of the ellipse described by the
electric field (polarization) does not depend on its size (intensity).

To proceed further we need to represent every component
$\op{\varrho}^{(S)}$ in a polarization basis. Instead of using
directly the states $\{ | S, m \rangle \}$, it is more convenient to
write such an expansion as
\begin{equation}
  \label{rho1}
  \op{\varrho}^{(S)} =  \sum_{K= 0}^{2S} \sum_{q=-K}^{K}  
  \varrho_{Kq}^{(S)}  \,   \op{T}_{Kq}^{(S)} \, ,
\end{equation}
where the irreducible tensor operators $\op{T}_{Kq}^{(S)}$
are~\cite{Varshalovich:1988ct}
\begin{equation}
  \label{Tensor} 
  \op{T}_{Kq}^{(S)} = \sqrt{\frac{2 K +1}{2 S +1}} 
  \sum_{m,  m^{\prime}= -S}^{S} C_{Sm, Kq}^{Sm^{\prime}} \, 
  |  S , m^\prime \rangle \langle S, m | \, ,
\end{equation}
with $ C_{Sm, Kq}^{Sm^{\prime}}$ being the Clebsch-Gordan coefficients
that couple a spin $S$ and a spin $K$ \mbox{($0 \le K \le 2S$)} to a
total spin $S$.  

Although at first sight Eq.~(\ref{Tensor}) might look a bit intricate,
$\op{T}^{(S)}_{Kq}$ is related to the $K$th power of the Stokes
operators, a simple observation that will turn out crucial in the
following. In particular, the monopole $\op{T}_{00}^{(S)}$, being
proportional to the identity, is always trivial, while the dipole
$\op{T}_{1q}^{(S)}$ is proportional to $\op{S}_{q}$ and thus
renders the classical picture, in which the state is depicted by its
average value. Therefore, higher-order multipoles embody the
polarization fluctuations we wish to appraise.

The expansion coefficients $ \varrho_{Kq}^{(S)} = \Tr [
\op{\varrho}^{(S)} \, T_{Kq}^{(S) \, \dagger} ]$ are known as state
multipoles, and they contain complete information, but sorted in a
manifestly su(2)-invariant form.

Alternatively, one can look at
\begin{equation}
  \label{eq:wk}
  \mathcal{W}_{K}^{(S)}  =   \sum_{q=- K}^{K} |\varrho_{Kq}^{(S)}|^{2} \, ,
\end{equation}
which is just the square of the state overlapping with the $K$th
multipole pattern in the $S$th subspace.  When there is a distribution
of photon numbers, we sum over all of them: $ \mathcal{W}_{K} =
\sum_{S} P_{S} \, \mathcal{W}_{K}^{(S)}$.  One can easily find out that
\begin{equation}
  \sum_{K} \mathcal{W}_{K} = 
  \Tr (  \op{\varrho}^{2} ) \, ,
\end{equation}
so it is just the purity.  Actually, as shown in the
Appendix~\ref{ap:A}, $\mathcal{W}_{K}$ can be interpreted as a measure
of the localization of the state in phase space.

\begin{table}[t]
  \caption{Values of $\mathcal{W}_{K}$ and the degree $\mathbb{P}_{K} $
    for three different quantum polarization states. $| S; \theta, \phi \rangle$ stands
    for an SU(2) coherent state in the $S$ subspace, and $|\alpha_{H},
    \alpha_{V}\rangle$ is a two-mode quadrature coherent state with
    $\bar{N} = |\alpha_{H} |^{2} + |\alpha_{V} |^{2}$ the average number
    of photons.}
  \label{table1}
  \begin{ruledtabular}
    \begin{tabular}{lcc}
      State  & $\mathcal{W}_{K}$  & $\mathbb{P}_{K}$ \\
      \hline
      $|S,  m\rangle$  & $\frac{2K+1}{2S+1}{(C_{Sm,K0}^{Sm})}^{2}$  & 
      $ \left [ 
     \frac{\sum_{\ell=1}^{K} \frac{2\ell+1}{2S+1}{(C_{Sm,\ell0}^{Sm})}^{2}}
      {\sum_{\ell=1}^{K}\frac{2\ell+1}{2S+1}{(C_{SS,\ell0}^{SS})}^{2}}
     \right ]^{1/2}$ \\
      $| S; \theta, \phi \rangle$  & $\frac{2K+1}{2S+1}{(C_{SS,K0}^{SS})}^{2}$
      & 1 \\
      $| \alpha_{H}, \alpha_{V} \rangle$  & $ \frac{2K+1}{2S+1} (
      C_{SS,K0}^{SS} ) ^{2}$  & $\sum_{S=K/2}^{\infty} 
      \frac{\bar{N}^{2S }e^{-\bar{N}}}{(2S)!}$
    \end{tabular}
  \end{ruledtabular}
\end{table}

\begin{figure*}
  \includegraphics[width=0.65\columnwidth]{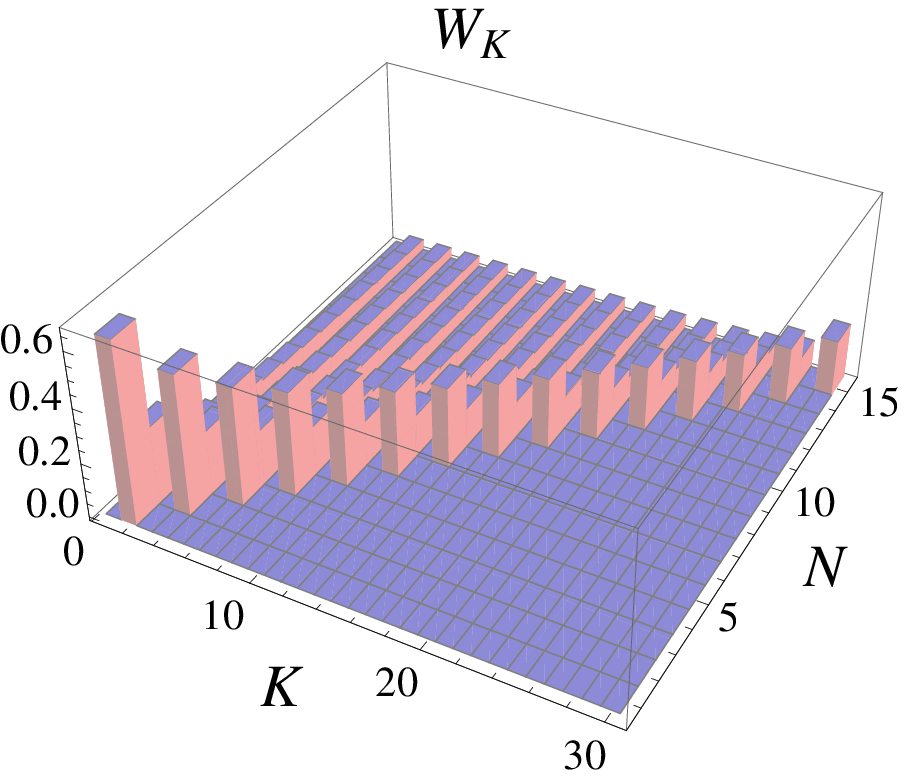}
  \quad
 \includegraphics[width=0.65\columnwidth]{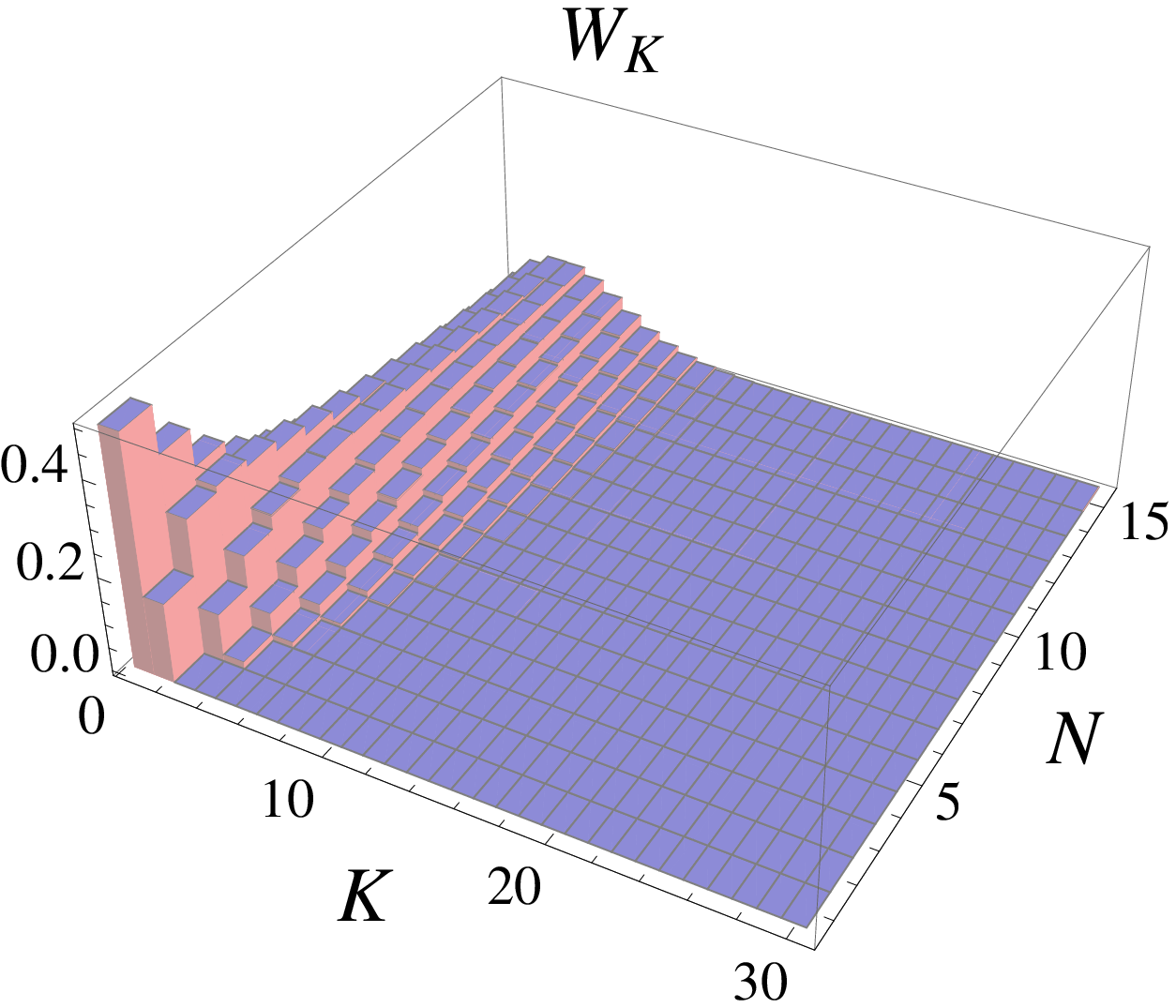}
\quad
  \includegraphics[width=0.65\columnwidth]{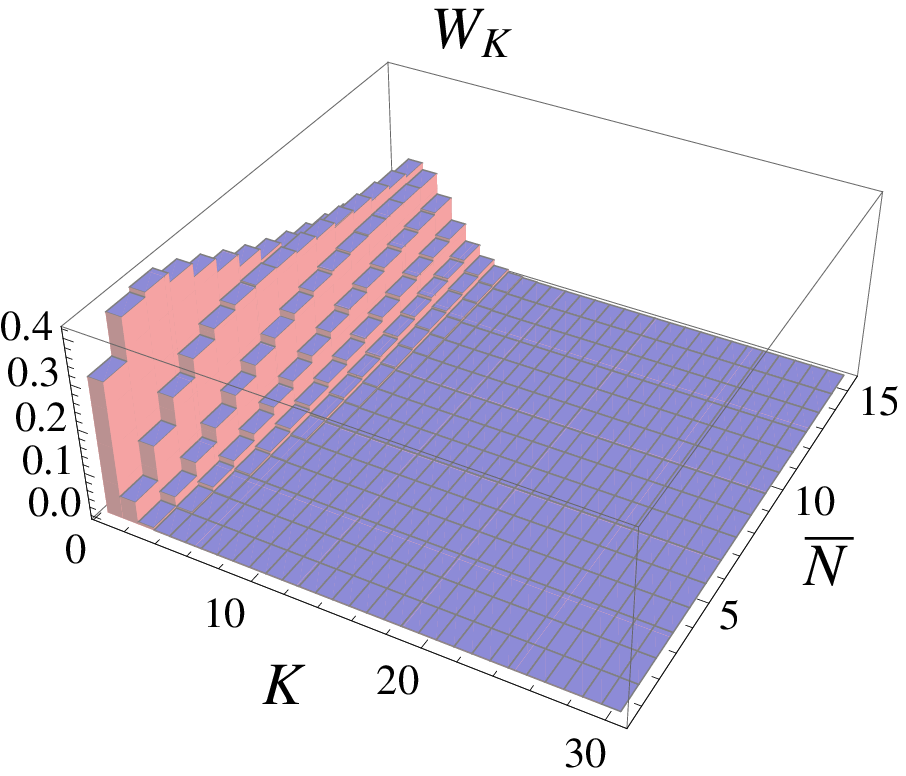}
  \caption{(Color online) Distribution $\mathcal{W}_{K}$ as a
    function of the multipole order $K$ for the state $|S, 0 \rangle $
    ($ |N, N \rangle$, with $S = N$, in the basis $|n_{H}, n_{V}
    \rangle$) and for a two-mode quadrature coherent state with
    average number of photons $\bar{N} = |\alpha_{H} |^{2} +
    |\alpha_{V} |^{2}$.}
  \label{fig:W_K}
\end{figure*}

In Table~\ref{table1} we have worked out the values of
$\mathcal{W}_{K}$ for three outstanding examples of quantum states
that will serve as a guide: the state $|S, 0 \rangle$ (which reads
$|N, N \rangle$, with $N = S$, in the basis $|n_{H}, n_{V} \rangle$),
the SU(2) coherent state $| S; \theta, \phi \rangle$ (defined in the
Appendix~\ref{ap:A}), and a two-mode quadrature coherent state $|
\alpha_{H}, \alpha_{V} \rangle$, summing up over the Poissonian
photon-number distribution (with $\bar{N} = |\alpha_{H}|^{2} +
|\alpha_{V}|^{2}$). In Fig.~\ref{fig:W_K} we also plot these cases in
point; as we can see, for the classical quadrature coherent state the
first multipoles contribute the most; whereas for the nonclassical
$|S, 0 \rangle$ state, the converse holds.

\section{Reconstructing the multipoles} 

The analysis thus far confirms that multipoles constitute a natural
tool to deal with polarization properties. We will show next that, in
addition, they can be experimentally determined.
 
\begin{figure}[b]
  \includegraphics[width=0.98\columnwidth]{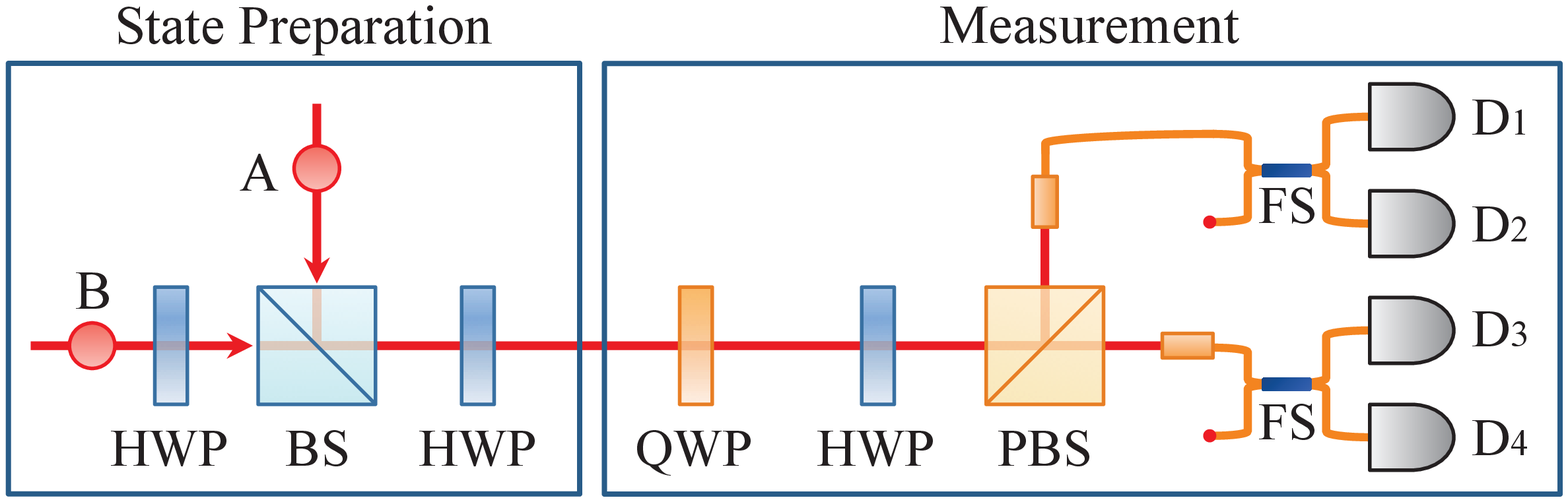}
  \caption{(Color online) Experimental setup. Single-photons A and B,
    both horizontally polarized, are prepared by spontaneous
    parametric down-conversion. (P)BS denotes a (polarization) beam
    splitter. HWP and QWP denote half-wave and quarter-wave plates,
    respectively. FS denotes a 50:50 fiber splitter and D1-D4 denote
    single-photon avalanche photodiodes.}
  \label{fig:setup}
\end{figure}

The polarization state is customarily analyzed with a Stokes
measurement setup (see Fig.~\ref{fig:setup}), consisting of a
half-wave plate (HWP) with the axis at angle $\theta$, followed by a
quarter-wave plate (QWP) at angle $\phi$, and a polarizing beam splitter
(PBS) that separates the $H$ and $V$ modes. The wave plates
effectively perform a displacement of the state that can be described
by the operator $\hat{D} (\theta, \phi) = e^{i \theta \op{S}_{2}} e^{i
  \phi \op{S}_{3}}$, and $(\theta, \phi)$ are angular coordinates on
the sphere. The two output of PBS are measured by photon detectors:
the photocurrent sum gives directly the eigenvalue of $\op{N}$, while
the difference gives the observable $ \op{S}_{\mathbf{n}} = \mathbf{n}
\cdot \op{\mathbf{S}}$, where $\mathbf{n}$ is the unit vector in the
direction~$(\theta, \phi)$~\cite{Marquardt:2007bh}.

Altogether, this indicates that the scheme yields the probability
distribution for $\op{S}_{\mathbf{n}}$, from which we can equivalently
infer the moments 
\begin{equation}
\mu_{\ell}^{(S)} (\theta, \phi ) = 
 \Tr [ \op{S}_{\mathbf{n}}^{\ell} \, \op{\varrho}^{(S)} ] \, . 
\end{equation}
For simplicity, we restrict ourselves to a subspace with fixed number
of photons $S$, but everything can be smoothly extended to the whole
polarization sector.

We start by noticing that the measurable moments can be expressed in
terms of the state multipoles as
\begin{eqnarray}
  \label{eq:lqh}
  \mu_{\ell}^{(S)} (\theta, \phi ) & = &  
  \Tr  [ \op{S}_{3}^{\ell} \,  \op{D} (\theta , \phi ) \, 
  \op{\varrho}^{(S)}  \,  \op{D}^\dagger  (\theta , \phi )  ] 
  \nonumber \\
  & = &  
  \Tr \left [   \hat{S}_{3}^{\ell} \, \sum_{K=0}^{2S} \sum_{q, q^{\prime}=-K}^{K} 
    \varrho_{Kq}^{(S)} \, D_{qq^\prime}^{K} (\theta , \phi ) \,
    \op{T}_{Kq}^{(S)} \right ] \,  ,
\end{eqnarray}
where $D_{m m^\prime}^{S} (\theta ,\phi ) = \langle S, m | \op{D}
(\theta ,\phi ) |S, m^\prime \rangle $ is the Wigner
$D$-function~\cite{Varshalovich:1988ct}.  To proceed further we need
to compute
\begin{eqnarray}
  \Tr [ \op{S}_{3}^{\ell} \, \op{T}_{Kq}^{(S)} ]  & = & 
  \delta _{q0}   \left [ \frac{S (S +1) (2 S+1)}{3} \right ]^{\ell/2}  
  \frac{3^{\ell/2} \sqrt{2 K +1}}{(2 S +1)^{( \ell +1)/2}} \nonumber \\
  & \times & \sum_{m}  \left ( C_{Sm,10}^{Sm} \right )^{\ell} 
  C_{Sm,K0}^{Sm} \, .
\end{eqnarray} 
Interestingly, we have that $C_{Sm,10}^{Sm}=m/\sqrt{S(S+1)}$ and
\begin{equation}
  \sum_{m} m^{\ell} C_{Sm,K0}^{Sm} = i^{\ell - K} 
  \left . \partial_{\omega}^{\ell} 
    \chi_{K}^{S}(\omega ) \right  |_{\omega=0} 
  \equiv  f_{K\ell}^{(S)} \, \qquad K \leq \ell \, ,
\end{equation}
with $\chi_{S}^{m}(\omega )$ the generalized
SU(2) character~\cite{Varshalovich:1988ct}.  Collecting all those
results together, the moments come out connected with the multipoles
in quite an elegant way:
\begin{equation}
  \label{eq:mrho}
  \mu_{\ell}^{(S)} (\theta, \phi ) = \sqrt{\frac{4\pi }{2 S + 1}}
  \sum_{K=0}^{\ell} \sum_{q=-K}^{K} \varrho_{Kq}^{(S)}  \,
  f_{K\ell}^{(S)}  \, Y_{Kq} (\theta, \phi )  \, ,
\end{equation}
$Y_{Kq} (\theta, \phi ) $ being the spherical harmonics.

We can benefit from the orthonormality of $Y_{Kq} (\theta, \phi)$ to
integrate Eq.~(\ref{eq:mrho}) so as to obtain
\begin{equation}
  \varrho_{Kq}^{(S)} = \frac{1}{ f_{K\ell}^{(S)}} 
  \sqrt{\frac{2 S +1}{4\pi}} \int_{\mathcal{S}^{2}} d\Omega \, 
  \mu_{\ell}^{(S)} ( \theta, \phi ) \, Y_{Kq}^{\ast} (\theta, \phi ) \, , 
\end{equation}
where $K \leq \ell$ and the integral extends over the whole unit
sphere $\mathcal{S}^{2}$ with $d\Omega= \sin \theta d\theta d\phi$
being the solid angle.  The reconstruction of the state requires the
knowledge of \textit{all} the multipoles: this implies measuring
\textit{all} the moments in \textit{all} the directions, which proves
to be very demanding~\cite{Muller:2012ys}.

Nonetheless, we can attack the problem in a much more economic way.
The central idea is that to determine the $L$th multipole, it is
enough to perform a Stokes measurement in $2L+1$ independent
directions. As a matter of fact, the proposal proceeds  in a recurrent way:  
first, we measure the first-order moments in the three coordinate axis
(or other equivalent ones) and reconstruct $ \varrho_{1q}^{(S)}$. That
is, from  the values of $\mu_{1}^{(S)} (\theta, \phi )$, which can
write down as,
\begin{equation}
\mu_{1}^{(S)} (\theta, \phi ) =f_{11}^{(S)} \sqrt{\frac{4\pi }{2S+1}} 
\sum_{q=-1}^{1} \varrho_{1q}^{(S)} Y_{1q}( \theta, \phi ) \, ,
\end{equation}
we need to know $\varrho_{1q}^{(S)}$. By taking into account that
$f_{11}^{(S)} = (2S +1) \sqrt{S (S +  1)}/3$, we can solve the
resulting linear system, getting 
\begin{equation}
  \left ( 
    \begin{array}{c}
      \varrho_{11}^{(S)} \\ 
      \varrho_{10}^{(S)}  \\ 
      \varrho_{1-1}^{(S)} 
    \end{array} 
  \right )  = 
  \sqrt{\frac{3}{2S (S+1)(2S+1)}} 
  \left ( 
    \begin{array}{ccc}
      -1 & i & 0 \\ 
      0 & 0 & \sqrt{2}  \\ 
      1 & i & 0
    \end{array}
  \right ) 
  \left ( 
    \begin{array}{c}
      \mu_{1,1}^{(S)}\\ 
      \mu_{1,2}^{(S)} \\ 
      \mu_{1,3}^{(S)}
    \end{array}
  \right )  \, ,  
  \label{S1_rec} 
\end{equation}
whence we infer all the first-order properties. Here $\mu_{1,k}^{(S)}$
indicate the first-order moment in the $k$th direction.

The measurement of the second moments gives us
\begin{equation}
 \mu_{2} (\theta, \phi) = \frac{1}{2 S + 1} f_{02}^{(S)} + 
f^{(S)}_{22} \sqrt{\frac{4\pi }{2S+1}} \sum_{q=-2}^{2} 
\varrho_{2q}^{(S)} Y_{2q} (\theta, \phi ) \, ,
\end{equation}
with
\begin{eqnarray}
f_{02}^{(S)} & = & \frac{1}{3}S(S+1)(2S+1) \, , \nonumber \\
& & \\ 
f_{22}^{(S)} & = & \frac{4(2S+1)}{5!}\sqrt{S(2S-1)(S+1)(2S+3)}\,,
\nonumber  
\end{eqnarray}
while $f_{12}^{(S)} =0$. We need to fix five optimal directions to
invert that system. For example, thinking of the measurements as
lines, we can choose the directions as
\begin{eqnarray}
\label{eq:ns}
  & 
  \mathbf{n}_{1,2} \propto 
  \left (
    \begin{array}{c}
      0  \\
      \pm 2 \\
      1 + \sqrt{5} 
    \end{array} 
  \right ) \, ,
  \qquad 
  \mathbf{n}_{3,4} 
  \propto 
  \left (
    \begin{array}{c}
      \pm 2 \\
      1 + \sqrt{5} \\
      0
    \end{array} 
  \right ) \, ,
  & \nonumber \\
  & \mathbf{n}_5 \propto
  \left (
    \begin{array}{c}
      1 + \sqrt{5} \\
      0 \\
      2
    \end{array} 
  \right ) \, , &
\end{eqnarray}
which maximizes the minimum angle between the lines and thus in some
sense spreads out the measurements over the Poincar\'{e} sphere as
much as possible~\cite{Conway:1996ys}.  The system can be then solved,
and all we need to characterize the process at second order is known.

For the $L$th moment, we have
\begin{equation}
  \bm{\mu}_{L}^{(S)} = \sqrt{\frac{4\pi }{2S+1}} f_{KL}^{(S)}
  \mathbf{Y}_{K} \,  \bm{\varrho}_{K}^{(S)} \, ,
\end{equation}
where $\bm{\mu}_{L}^{(S)} = \left (\mu_{L} (\theta_{1}, \phi_{1}) ,
  \ldots, \mu_{L} (\theta_{2L+ 1}, \phi_{2L + 1}) \right )$, and
similarly for $\bm{\varrho}_{K}$ and $[ \mathbf{Y}_{L} ]_{ij} = Y_{Lj
}(\theta_{i}, \phi_{i})$. Observe that, in general, the right-hand
side hinges on the results of lowest-order measurements. The linear
inversion of that equation can be formally written down as
\begin{equation}
  \bm{\varrho}^{(S)}_{K} = \frac{1}{f_{KL}^{(S)}}
  \sqrt{\frac{2S+1}{4\pi}}\frac{4\pi }{2L+1}
  \mathbf{P}_{L}^{-1} \mathbf{Y}_{L}^{\dagger } \bm{\mu}_{L}^{(S)} \,   , 
\end{equation}
where $\mathbf{P}_{L} =4 \pi/(2L+1)
\mathbf{Y}_{L}\mathbf{Y}_{L}^{\dagger}$, with $[ \mathbf{P}_{L} ]
_{ij}= P_{L}(\omega _{ij})$,  $\cos \omega _{ij} =\cos \theta
_{i}\cos \theta _{j}+\sin \theta _{i}\sin \theta _{j}\sin (\phi
_{i}-\phi _{j})$ and $P_{L}(\omega _{ij})$ is the Legendre polynomial.
The choosing of the appropriate directions is, in general, a tricky
question if one wants to be sure about the linear independence, but it 
has been thoroughly studied~\cite{Filippov:2010kx}. In practice,
methods such as maximum likelihood are much more efficient in handling that
inversion~\cite{ML:2004kl}.

To check the proposed strategy, we have performed an experiment using
spontaneous parametric down-conversion. The photon pairs centered at
780 nm were generated in a 2~mm thick type-I $\beta$-barium-borate
(BBO) crystal pumped by a femtosecond laser pulse centered at 390 nm
and subsequently filtered by an interference filter with a 4~nm
bandwidth and brought to the inputs of a Hong-Ou-Mandel
interferometer.  After the interferometer, either the state
$|1_{H},1_{V} \rangle$ or the state $| 2_{H}, 0_{V} \rangle$ can be
postselected, depending on the relative polarizations of the incident
photons.

\begin{table*}
\caption{\label{tab:table2}
Experimental and theoretical results obtained for the state $|2_{H},
0_{V} \rangle$ (which is the $|1,1 \rangle$ state in the angular momentum
basis). The number in parentheses indicate the error in the last 
figure. The directions of measurement are the three coordinate axes
for $\mu_{1}$ and (\ref{eq:ns}) for $\mu_{2}$.}
\begin{ruledtabular}
\begin{tabular}{cccccccccc}
 &\multicolumn{2}{c}{Experiment} &\multicolumn{2}{c}{Theory} & 
 & \multicolumn{2}{c}{Experiment}  &\multicolumn{2}{c}{Theory} \\
Direction & $\mu_{1}$ & $\mu_{2}$ & $\mu_{1}$ & $\mu_{2}$ & Multipole &
$K=1$ & $K=2$ & $K=1$ & $K=2$ \\ \hline
1 & $- $ 0.10 (3)  &  0.84 (7)  & 0 & 0.8618  & 
$\varrho_{K-2}$ &  & $-$ 0.01 (7) $ -$ 0.01 (1) $i$  & & $0$ \\
2  & 0.06 (2)  & 0.87 (1) & 0 &  0.8618
& $\rho_{K-1}$ & $-$ 0.05 (2) + 0.03 (1)$ i$ &  0.07 (6) $ - $ 0.02 (2) $i $ & 0 & 0 \\
3 & 0.99 (3)  &  0.50 (2)  & 1 &  0.5000 &
$\rho_{K0}$ & 0.70 (1) &  0.39 (4)  & 0.7071 & 0.4082  \\
4 &  & 0.52 (1) & & 0.5000 & 
$\rho_{K1}$ & 0.05 (2) + 0.03 (1) $ i $ &  $-$ 0.07 (7) $- $ 0.02 (1) $ i$ & 0 & 0 \\
5 & & 0.70 (2) & & 0.6382 & 
$\rho_{K2}$ & & $-$ 0.01 (1) +0.01 (1) $i$  &  & 0
\end{tabular}
\end{ruledtabular}
\end{table*}

The setup is sketched in Fig.~\ref{fig:setup}. At each output of the
PBS, a two-photon detector is simulated by a 50:50 fiber beam splitter
(FS) and two single-photon detectors (PerkinElmer, SPCM-AQRH).  The
photon detection efficiency of each single-photon detector channel is
used to calibrate the measurement of the Stokes parameters. To achieve
full information about the first and second order moments, we have
measured these coincidences in five distinct measurement bases and
then reconstructed the multipoles via linear inversion.  Each
measurement is done for 3~s and repeated three times to improve the
precision.

In Table~\ref{tab:table2} we summarize the results obtained for the
state $|2_{H}, 0_{V} \rangle$ (which is $|1 , 1 \rangle$ in the
angular momentum basis). The agreement with the theory is pretty
good. Although this instance might look a bit naive, it constitutes
quite a conclusive proof of principle of our method.

\section{Assessing higher-order polarization correlations} 

Even though  the polarization information is encoded in the set $\{
\mathcal{W}_{K}^{(S)} \}$, for most of the states only a limited
number of multipoles play a substantive role and the rest of them have
an exceedingly small contribution, so that gaining a good feeling
of the corresponding behavior may be tricky.

\begin{figure}[b]
  \includegraphics[width=0.45\columnwidth]{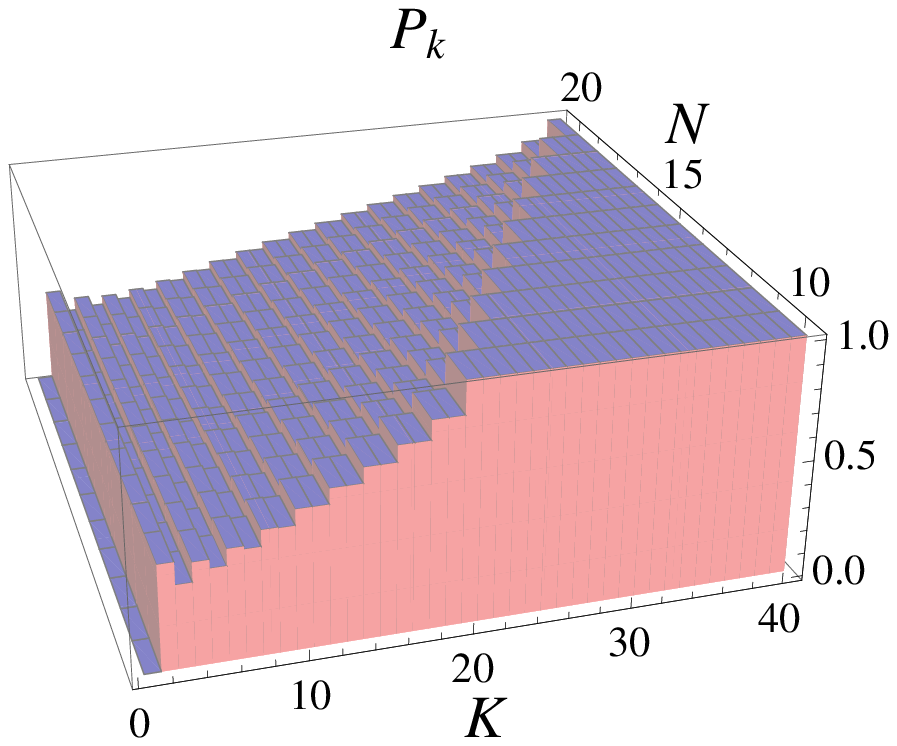}
  \qquad
  \includegraphics[width=0.45\columnwidth]{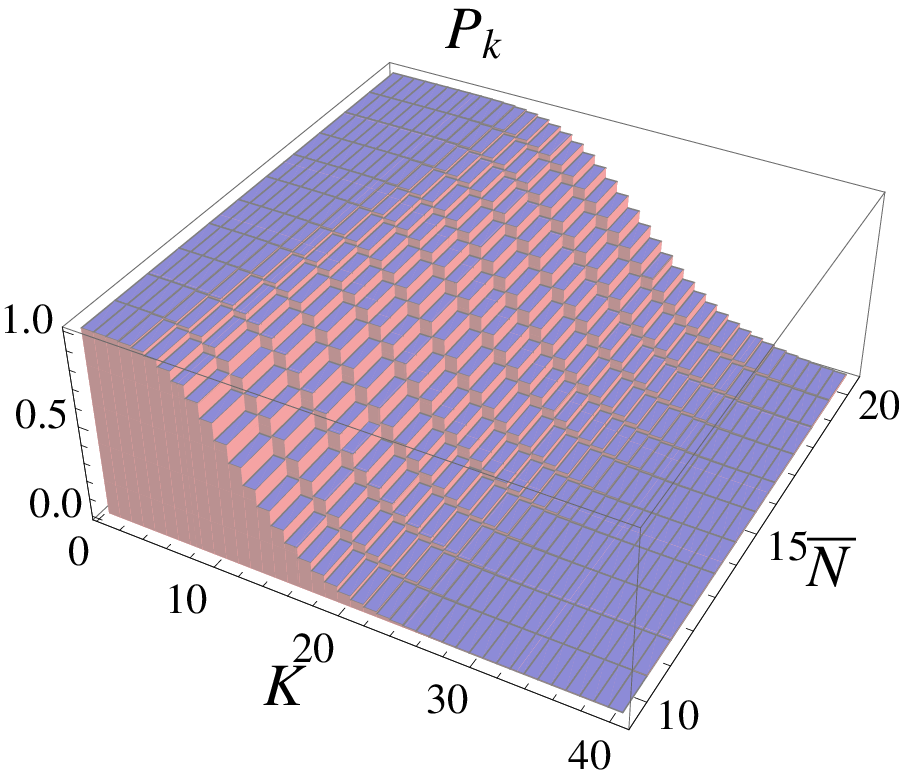}
  \caption{(Color online) Degree of polarization $\mathbb{P}_{K}$ as a
    function of the multipole order $K$ for the state $|S, 0 \rangle$
    (left panel) and a quadrature coherent state $| \alpha_{H}, \alpha_{V}
    \rangle$ with
    average number of photons $\bar{N} = |\alpha_{H} |^{2} +
    |\alpha_{V} |^{2}$ (right panel).  
    \label{fig:PK}}
\end{figure}

A possible way to bypass this disadvantage is to look at the
cumulative distribution
\begin{equation}
  \label{eq:cum}
  \mathcal{A}^{(S)}_{K} = \sum_{\ell = 1}^{K} \mathcal{W}^{(S)}_{K} \, ,
\end{equation}
which conveys the whole information \textit{up} to order $K$. We know
from probability that it has remarkable
properties~\cite{Jaynes:2003uq}. Moreover, our previous reconstruction
puts in clear evidence that to obtain the $K$th multipole, one needs
to determine all the previous moments.

\begin{figure}[b]
  \includegraphics[width=0.47\columnwidth]{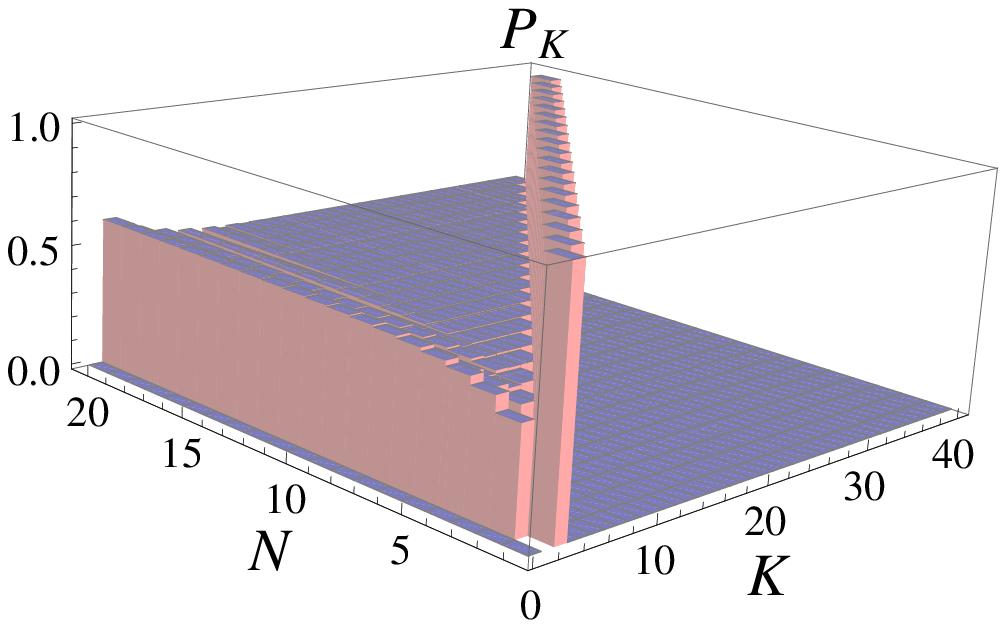}
  \quad
  \includegraphics[width=0.47\columnwidth]{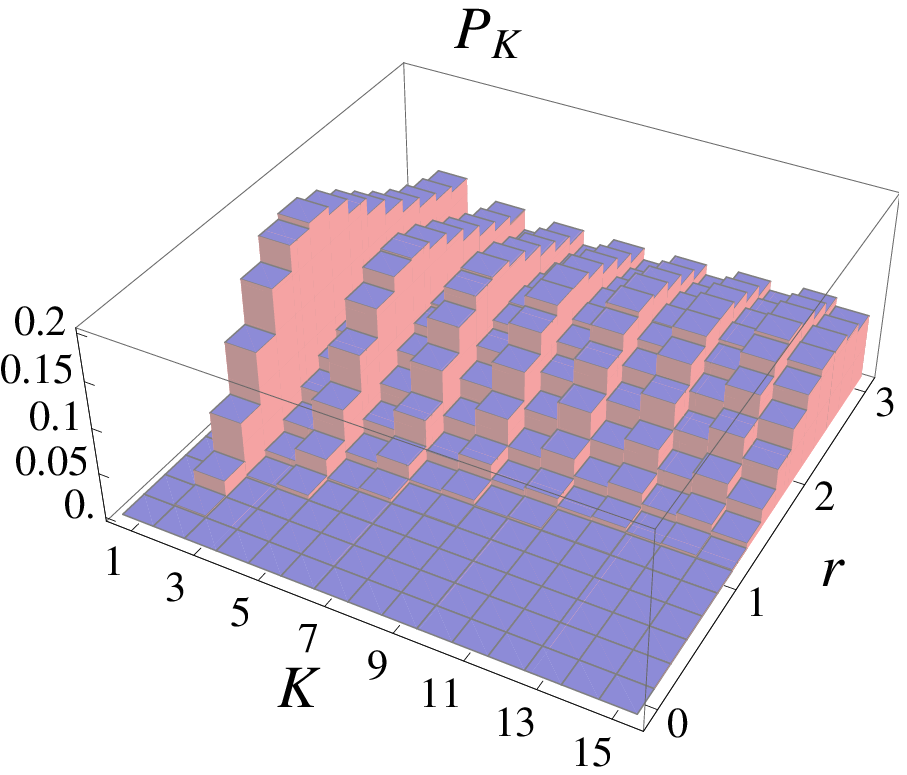}
  \caption{(Color online) Degree of polarization $\mathbb{P}_{K}$ for
    the NOON (left panel) and two-mode squeezed vacuum (right panel)
    using the squeezing parameter $r$ as a measure of the average
    number of photons.}
  \label{fig:W_Kex}
\end{figure}

As with any cumulative distribution, $ \mathcal{A}^{(S)}_{K} $ is a
monotone nondecreasing function of the multipole order, with
$\mathcal{A}^{(S)}_{2S}$ being proportional to the state purity
[except by the monopole contribution, $K=0$, which is not included in
(\ref{eq:cum})]. One might be interested in dealing instead with
magnitudes satisfying $0 \le \mathbb{P}_{K} \le 1$ for every $K$, as
any sensible degree of polarization~\cite{Bjork:2010rt}. To that end,
we note that for SU(2) coherent states we have
\begin{equation}
  \label{eq:Aksu2}
  \mathcal{A}^{(S)}_{K,\mathrm{SU(2)}} = \frac{2S}{2S +1} -
  \frac{[\Gamma (2S + 1)]^{2}}{\Gamma (2S-K) \Gamma (2S + K +2)} \, .
\end{equation}
After many numerical experiments, we conjecture that
$\mathcal{A}^{(S)}_{K,\mathrm{SU(2)}} $ is indeed maximal for any $K$
in each subspace $S$. This seems to suggest a degree of polarization
up to the $K$th order as
\begin{equation}
  \label{eq:PK}
  \mathbb{P}_{K} =
  \sum_{S} P_{S} \, \sqrt{\frac{\mathcal{A}^{(S)}_{K}}{\mathcal{A}^{(S)}_{K,\mathrm{SU(2)}}}} \, .
\end{equation}
According to the definition (\ref{eq:PK}), $\mathbb{P}_{K}= 1$ (for
every $K$) for SU(2) coherent states, which is compatible with the idea
that they are the most localized states over the sphere. On the other
hand, for quadrature coherent states, which constitute an acid test
for any new proposal in polarization, the result, as indicated in
Table~I, reads
\begin{equation}
  \mathbb{P}_{K} = \sum _{S=K/2}^{\infty } 
  \frac{e^{-\bar{N}}\bar{N}^{2S}}{(2S)!} \simeq 
  \frac{1}{2} \text{erfc} 
  \left(\frac{K-\bar{N}}{\sqrt{2 \bar{N}}} \right ) \, . 
\end{equation}
Here, $\bar{N}$ is the average number of photons and the second
equality, in terms of the complementary error function, holds true for
$\bar{N} \gg 1$. From the properties of this function, we can estimate
that the multipoles that contribute effectively are, roughly speaking,
from 1 to $\bar{N}$. In Fig.~\ref{fig:PK} we plot $\mathbb{P}_{K}$ for
the states $|S, 0 \rangle$ and $| \alpha_{H}, \alpha_{V} \rangle$.

To round off our understanding of $\mathbb{P}_{K}$, in
Fig.~\ref{fig:W_Kex} we have depicted $\mathbb{P}_{K}$ for  two
other relevant quantum states routinely treated in this context; NOON
and two-mode squeezed vacuum states, defined as
\begin{eqnarray}
  \label{eq:3}
  |\mathrm{NOON} \rangle & = & \frac{1}{\sqrt{2}}
  ( |N, 0 \rangle + |0, N   \rangle ) \, , \nonumber \\
  &  & \\
  |\mathrm{TMSV} \rangle & = & \sqrt{1- \lambda^{2}} \sum_{N}
  \lambda^{N} |N, N \rangle  \, . \nonumber 
\end{eqnarray}
To follow the standard notation, in both cases we have employed the
$\{ | n_{H}, n_{V} \rangle \} $ basis and $\lambda = \tanh r$, with $r$
the squeezing parameter. 

For the particular yet significant case of the dipole ($K=1$),
Eq.~(\ref{eq:PK}) reduces to 
\begin{equation}
  \mathbb{P}_{1}= \sum_{S}^\infty P_{S}  
  \frac{ \sqrt{\langle \op{S}_{1} \rangle^2+ \langle \op{S}_{2} \rangle^2 +
    \langle \op{S}_{3} \rangle^2}}{\langle \op{S}_{0} \rangle} \, ,
\end{equation}
and the average values are calculated in every subspace
$S$. Interestingly, this definition has been recently proposed as a
way to circumvent the shortcomings of the standard degree of
polarization~\cite{Kothe:2013fk}; in our approach, it emerges quite in
a natural way.

To close our paper, we briefly consider the instance of
$\mathbb{P}_{2}$. For two-mode quadrature coherent states
$|\alpha_{H}, \alpha_{V} \rangle$ we immediately get
\begin{equation}
  \mathbb{P}_{2} (|\alpha_{H}, \alpha_{V} \rangle ) =
  1 -  (1+ \bar{N}) \exp (- \bar{N}) \, ,
\end{equation}
which tends to the unity when the average number of photons $\bar{N}$
becomes large enough, in agreement with previous second-order
approaches~\cite{Klimov:2010uq}. For the states $| S, m\rangle $, we
have
\begin{equation}
  \mathbb{P}_{2} (| S, m \rangle ) = 
  \frac{45 m^4+5 S^2 (S + 1)^2 - 9 m^2 [2 S (S+1) + 1]}
  {4 S^2  (2 S-1) (4 S + 1)} \, . 
\end{equation}
This expression is exactly unity whenever $ m = \pm S$ or $ m = \pm
\sqrt{1+2 S-3 S^2}/\sqrt{5}$ (this equality is valid only for $m$
integer).  This latter condition is only met when $ S=1$ with
$m=0$.

\begin{figure}[t]
  \includegraphics[width=0.75\columnwidth]{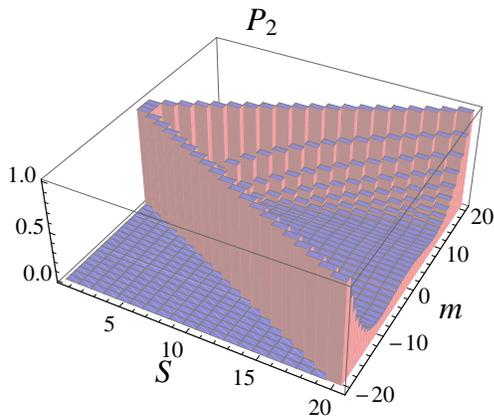}
  \caption{(Color online) Second-order degree of polarization
    $\mathbb{P}_{2}$ for the state $|S, m \rangle$.}
  \label{fig:P2_SM}
\end{figure}

On the other hand, $\mathbb{P}_{2}$ attains its minimum value 
\begin{equation}
  \label{eq:min}
    \mathbb{P}_{2,\mathrm{min}} (| S, m \rangle )  = 
\frac{9+18 S+8 S^2}{80 S^2} \simeq \frac{1}{10} \, ,
\end{equation}
whenever $m = \pm \sqrt{1+2 S+2 S^2}/\sqrt{10}$.
In Fig.~\ref{fig:P2_SM} we outline these facts.

\section{Concluding remarks} 

Multipolar expansions are a commonplace and a formidable tool in many
branches of physics.  We have applied such an expansion to the
polarization density matrix, showing how the corresponding state
multipoles quantify higher-order fluctuations in the Stokes
variables. In this way we have provided the first systematic
characterization of quantum polarization fluctuations that,
paradoxically, was missing in the realm of quantum optics.

 Moreover,  the formalism can be manifestly extended to other systems
 in which SU(2) symmetry plays a crucial role (such as in
 Bose-Einstein condensates, spin chains, etc) and to other unitary
 symmetries, such as SU(3) (which is pivotal to understanding the
 polarization properties of the near field). This is more than
 an academic curiosity, and  work in this direction is ongoing in our group.

\begin{acknowledgments}
  Financial support from the Swedish Foundation for International
  Cooperation in Research and Higher Education (STINT), the Swedish
  Research Council (VR) through its Linn{\ae}us Center of Excellence
  ADOPT and contract No.~621-2011-4575, the CONACyT (Grant
  No.~106525), the EU FP7 (Grant Q-ESSENCE), and the Spanish DGI
  (Grant FIS2011-26786) is gratefully acknowledged. It is also a
  pleasure to thank H. de Guise for stimulating discussions.
\end{acknowledgments}

\appendix

\section{Polarization quasidistributions}
\label{ap:A}

The discussion in this paper suggests that polarization must be specified by
a probability distribution of polarization states. As a matter of fact,
such a probabilistic description is unavoidable in quantum optics from
the very beginning, since $\{ \op{S}_{k} \}$ do not commute and thus
no state can have a definite value of all of them simultaneously.

The SU(2) symmetry inherent in the polarization structure of quantum
fields allows us to take advantage of the pioneering work of
Stratonovich~\cite{Stratonovich:1956qc} and
Berezin~\cite{Berezin:1975mw}, who worked out quasiprobability
distributions on the sphere satisfying all the pertinent requirements.
This construction was later generalized by
others~\cite{Agarwal:1981bd,Brif:1998if,Heiss:2000kc,
  Klimov:2000zv,Klimov:2008yb} and has proved to be very useful in
visualizing properties of spinlike
systems~\cite{Dowling:1994sw,Chumakov:1999sj,Klimov:2002cr}.

For each partial $\op{\varrho}^{(S)}$, one can define 
$r$-parametrized SU(2) quasidistributions as
\begin{equation}
  \label{eq:QSU2j}
  W_{r}^{(S)} (\theta, \phi) =  \frac{\sqrt{4 \pi}}{\sqrt{2S+1}} 
  \sum_{K=0}^{2S} \sum_{q=-K}^{K} (C_{SS,K0}^{SS})^{-r} \, \varrho_{Kq}^{(S)} \,
  Y_{Kq}^{\ast} (\theta, \phi) \, .   
\end{equation}
For $r=0$ this is the Wigner function, while $r = + 1$ and $-1$ leads to the
$P$ and $Q$ functions, respectively. Note also that the Clebsch-Gordan
coefficient $C_{SS,K0}^{SS}$ has a very simple analytical
form~\cite{Varshalovich:1988ct}:
\begin{equation}
  \label{eq:Cesp}
  C_{SS,K0}^{SS} = \frac{\sqrt{2S+1} (2S)!}
  {\sqrt{(2S-K)! \, (2S+1 + K)!}} \, .
\end{equation}
While, for spins, $S$ is typically a fixed number, in quantum optics
most of the states involve a full polarization sector and one should
sum over the subspaces contributing to the state.

The integral
\begin{equation}
  \label{eq:loc}
  \Sigma = \frac{1}{\int d\Omega \, 
    [W^{(S)}_{r} (\theta, \phi)  ]^{2}} \, ,
\end{equation}
extended to the whole sphere, can be interpreted as the effective area
where the corresponding quasidistribution is different from zero. In
other words, $\Sigma$ is a measure of the number of polarization
states contained in a given field state. This and similar definitions
have already been used as measures of localization and uncertainty in
different contexts~\cite{Hall:1999oq}.

Using the explicit form of (\ref{eq:QSU2j}) we immediately get
\begin{equation}
  \label{eq:2}
  \int d\Omega \,  [W^{(S)}_{r} (\theta, \phi) ]^{2} = 
  \frac{4 \pi}{2S+1} 
  \sum_{K=0}^{2S}  (C_{SS,K0}^{SS})^{-2r} \mathcal{W}_{K}^{(S)}
  \, .
\end{equation}
We can appreciate a deep connection (except for the unessential
Clebsch-Gordan coefficient) between the distribution $\{
\mathcal{W}_{K}^{(S)} \}$ and the notion of localization in phase
space. In particular, for the Wigner function $r=0$ and the right-hand
side of (\ref{eq:2}) is giving information about the measured $\{
\mathcal{W}_{K}^{(S)} \}$.

For the sake of completeness, we briefly recall the definition of
the SU(2) coherent states (also known as spin or atomic coherent
states), which reads~\cite{Arecchi:1972zr,Perelomov:1986ly}
\begin{equation}
  \label{eq:defCS}
  |S; \theta , \phi \rangle = \op{D} (\theta, \phi ) 
  |S, -S \rangle \,  .
\end{equation}
Here $\op{D} (\theta, \phi ) = \exp (\xi \op{S}_{+} - \xi^{\ast}
\op{S}_{-})$ [with $\xi = (\theta / 2) \exp (- i \phi)$ and $(\theta,
\phi)$ being spherical angular coordinates] plays the role of a
displacement on the Poincar\'e sphere of radius $S$.

The ladder operators $\op{S}_{\pm} = \op{S}_{1} \pm i \op{S}_{2}$
select the fiducial state $|S, -S \rangle$ as usual: $\op{S}_{-} |S ,
- S \rangle = 0$.  This definition closely mimics its standard
counterpart for position and momentum.

Note that these coherent states are eigenstates of the measured
operator  $\op{S}_{\mathbf{n}} = \mathbf{n} \cdot \op{\mathbf{S}}$
\begin{equation}
  \label{eq:CSe}
  \op{S}_{\mathbf{n}}  |S; \theta , \phi \rangle = S |S; \theta , \phi
  \rangle \, ,
\end{equation}
and they saturate the uncertainty relation (\ref{eq:unrel}), so they
are the minimum uncertainty states  in polarization optics.
 
The two-mode quadrature coherent states $|\alpha_{H}, \alpha_{V}
\rangle$ can be expressed as a Poissonian superposition of SU(2)
coherent states
\begin{equation}
  |\alpha_{H}, \alpha_{V} \rangle = \exp(- \bar{N}/2) \sum_{S}
  \frac{\bar{N}^{2S }e^{-\bar{N}}}{(2S)!} |2S, \theta, \phi \rangle \, , 
\end{equation}
where $\bar{N} = |\alpha_{H}|^{2} + |\alpha_{V}|^{2}$ is the average
number of photons. 


%

\end{document}